\documentclass[sigconf,screen,nonacm]{acmart}

\usepackage{multirow}
\usepackage{listings}
\usepackage{xcolor}
\usepackage{paralist}
\usepackage{booktabs}
\usepackage{ulem}
\usepackage[
  framemethod=TikZ,
  backgroundcolor=backcolor, 
  roundcorner=8pt,
  leftmargin=1, 
  rightmargin=1, 
  innerleftmargin=5, 
  innertopmargin=5,
  innerbottommargin=5, 
  outerlinewidth=1, 
  linecolor=backcolor
]{mdframed}

\mdfdefinestyle{compact}{
  framemethod=TikZ,
  backgroundcolor=backcolor, 
  roundcorner=8pt,
  leftmargin=1, 
  rightmargin=1, 
  innerleftmargin=0, 
  innerrightmargin=0, 
  innertopmargin=0,
  innerbottommargin=0, 
  outerlinewidth=1, 
  linecolor=backcolor
}

\usepackage[htt]{hyphenat}

\definecolor{code}{rgb}{0.1,0.2,0.1}
\definecolor{codegray}{rgb}{0.5,0.5,0.5}
\definecolor{codepurple}{rgb}{0.58,0,0.82}
\definecolor{backcolor}{rgb}{0.96,0.96,0.96}
\definecolor{lightblue}{rgb}{0.9,0.9,1}

\lstdefinestyle{lststyle}{
    commentstyle=\color{code},
    keywordstyle=\color{magenta},
    numberstyle=\tiny\color{lightblue},
    stringstyle=\color{orange},
    basicstyle=\ttfamily\footnotesize,
    breakatwhitespace=false,
    breaklines=true,
    captionpos=b,
    keepspaces=true,
    numbersep=5pt,
    showspaces=false,
    showstringspaces=false,
    showtabs=false,
    tabsize=2,
}

\AtBeginDocument{\providecommand\BibTeX{{\normalfont B\kern-0.5em{\scshape i\kern-0.25em b}\kern-0.8em\TeX}}}

\setcopyright{none}
\copyrightyear{2018}
\acmYear{2018}
\acmDOI{XXXXXXX.XXXXXXX}

\acmConference[Conference acronym 'XX]{Make sure to enter the correct
  conference title from your rights confirmation email}{June 03--05,
  2018}{Woodstock, NY}

\begin{document}

\title{Enhancing Security of AI-Based Code Synthesis with GitHub Copilot via Prompt-Engineering}
\title{Enhancing Security of AI-Based Code Synthesis with GitHub Copilot via Cheap and Efficient Prompt-Engineering}

\author{Jakub Res}
\email{iresj@fit.vut.cz}
\orcid{0009-0004-6055-5136}
\affiliation{\institution{Brno University of Technology, Faculty of Information Technology}
\country{Czech Republic}
}

\author{Ivan Homoliak}
\email{ihomoliak@fit.vut.cz}
\orcid{0000-0002-0790-0875}
\affiliation{\institution{Brno University of Technology, Faculty of Information Technology}
\country{Czech Republic}
}

\author{Martin Pere{\v{s}}{\'\i}ni}
\email{iperesini@fit.vut.cz}
\orcid{0000-0002-2875-9567}
\affiliation{\institution{Brno University of Technology, Faculty of Information Technology}
\country{Czech Republic}
}

\author{Ale\v{s} Smr\v{c}ka}
\email{smrcka@fit.vut.cz}
\orcid{0009-0003-9535-5962}
\affiliation{\institution{Brno University of Technology, Faculty of Information Technology}
\country{Czech Republic}
}

\author{Kamil Malinka}
\email{malinka@fit.vut.cz}
\orcid{0000-0002-9009-2193}
\affiliation{\institution{Brno University of Technology, Faculty of Information Technology}
\country{Czech Republic}
}

\author{Petr Hanacek}
\email{hanacek@fit.vut.cz}
\orcid{0000-0001-5507-0768}
\affiliation{\institution{Brno University of Technology, Faculty of Information Technology}
\country{Czech Republic}
}

\renewcommand{\shortauthors}{Jakub Res, et al. }

\begin{abstract}
AI assistants for coding are on the rise. 
  However one of the reasons developers and companies avoid harnessing their full potential is the questionable security of the generated code. 
  This paper first reviews the current state-of-the-art and identifies areas for improvement on this issue. 
  Then, we propose a systematic approach based on 
  prompt-altering methods to achieve better code security of (even proprietary black-box) AI-based code generators such as GitHub Copilot, while minimizing the complexity of the application from the user point-of-view, the computational resources, and operational costs. 
  In sum, we propose and evaluate three prompt altering methods: (1) scenario-specific, (2) iterative, and (3) general clause, while we discuss their combination. 
  Contrary to the audit of code security, the latter two of the proposed methods require no expert knowledge from the user.
  We assess the effectiveness of the proposed methods on the GitHub Copilot using the OpenVPN project in realistic scenarios, and we demonstrate that the proposed methods reduce the number of insecure generated code samples by up to 16\% and increase the number of secure code by up to 8\%. 
  Since our approach does not require access to the internals of the AI models, it can be in general applied to any AI-based code synthesizer, not only GitHub Copilot.
\end{abstract}

\maketitle

\section{Introduction}

With the release of ChatGPT~\cite{ChatGPT} , public attention shifted towards AI assistant tools. 
These assistants are proficient in many areas, including software engineering or coding. The advent of AI coding assistants means transitioning from intelligent code-completion tools to code-generating tools. 
Although these AI assistants are far from perfect, in terms of solving coding problems, a recent model AlphaCode 2, proposed by Deepmind, scored better than over 85 \% of human competitors~\cite{alphacode2}. 

According to Liang et al.~\cite{usability_survey}  in the survey with 410 Github users' responses, 70~\% of respondents who had experiences with Github Copilot utilize it at least once in a month while 46~\% utilize the AI assistant daily. 
The most frequent reasons for developers using AI assistants were fewer keystrokes to write code and faster coding. 
Due to the rapidly rising popularity of AI assistants, researchers started to focus on studying the quality of the synthesized code and ways of improving it (see \autoref{code_quality}). 
While observing the validity or correctness, many studies overlook the crucial aspect of code---security.

In the motivating example, the AI assistant was tasked with generating a code snippet to fill a gap in the context of a C program.
Its objective was to create a new instance of the structure "\texttt{person}" and assign a~status value of zero to it.
Although the AI assistant provided a reasonable code (see \autoref{fig:example}), the snippet
contain CWE-476~\cite{CWE_476} (the \texttt{malloc} function could fail to allocate memory, thus resulting in a NULL pointer dereference).

\begin{figure}
\begin{mdframed}
\begin{lstlisting}[language=C, style=lststyle]
person *newPerson = (person *)malloc(sizeof(person));
newPerson->status = 0;
\end{lstlisting}
\end{mdframed}
\caption{Example of security issue generated by AI. The scenario comes from the dataset proposed in~\cite{asleepatkeyboard}. }
\label{fig:example}
\end{figure}

In this research, we aim to study various ways of improving code security generated by any proprietary Large Language Models (LLMs), and
we demonstrate our approach on the well-known GitHub Copilot~\cite{github_copilot}.

There exist a few categories for improving the code synthesis of AI models, such as output optimization, model fine-tuning, and prompt engineering, and each of them has some pros and cons.
In this work, we focus on efficiency, generality, and low costs, and therefore prompt engineering is the most suitable technique for us.
While literature for prompt engineering is mostly general~\cite{OpenAI_Prompt_Engineering}\cite{white2023prompt}\cite{s10439-023-03272-4}\cite{10.1145/3545945.3569823}, we are more specific and determine four approaches to it, which we further investigate:
\begin{inparaenum}
	\item \textbf{scenario-specific} information and warning providing,
	\item \textbf{iterative} security-specific prompting, 
	\item \textbf{general alignment shifting} using inception prompt (i.e., general clause), 
	\item \textbf{cooperative agents} system. \end{inparaenum}
In particular, we experiment with the former three approaches that are orthogonal in their principles.

\paragraph{\textbf{Contributions}}
The contributions of our paper are as follows:
\begin{enumerate}
\item We reviewed the literature and identified three different areas of code synthesis improvements of LLMs, involving \textit{optimizing the output}, \textit{model fine-tuning}, and \textit{prompt optimizations}.
    \item With the focus on generality, speed, and low costs, we aimed at prompt engineering area, and we proposed a systematic approach to enhancing its generated code security with three methods and their combinations.
    \item We evaluated the efficiency of proposed methods for prompt alteration on a real-world project OpenVPN and we managed to increase the ratio of secure code generated by up to 8\% and decrease the ratio of generated insecure code by up to 16\%.
\end{enumerate}

\paragraph{\textbf{Organization}}
In \autoref{sec:idea} we define the important terms for our paper and set a design space.
In \autoref{sec:proposed} we describe the proposed methods of prompt improvement.
In \autoref{sec:experiments} we describe the design of the experiment, methodology, dataset, and assessment of security with measured results.
We refer to the related work in \autoref{sec:related}.
We discuss the limitations and areas for future research in \autoref{sec:discussion}.
In \autoref{sec:conclusion} we conclude our work.
 \section{Background and design space}\label{sec:idea}\subsubsection*{\textbf{Prompt.}}
The prompt, in the context of this work, refers to the tuple: (1) \textbf{a task} that contains function declaration and its description, (2) code of the \textbf{context}, and (3) the user-specified \textbf{code commentary} 
related to security.

\subsubsection*{\textbf{Improvements of Code Synthesis.}}
In general, the literature contains three main areas of possible improvements to the LLM code-generating abilities (see \autoref{fig:optimizers}):

\begin{figure}[b]
\centering
\includegraphics[width=.9\linewidth]{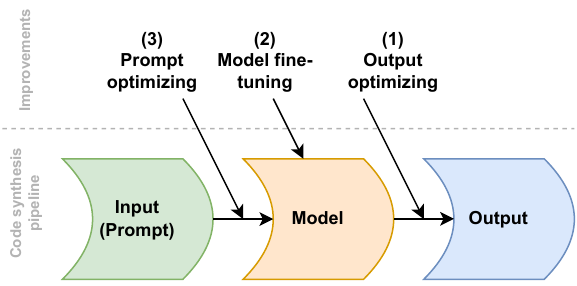}
\caption{Potential improvements of code synthesis. }
\label{fig:optimizers}
\end{figure}
\begin{enumerate}
\item{\textbf{Output optimizing}}
-- The first and the most intuitive approach is to post-process the output. 
Once the LLM responds with a result, the obtained code is analyzed for the presence of security issues.
Although the output correction is addressed by many works~\cite{ugare2024improving}\cite{wang2023factcheckgpt}\cite{vernikos2023small}, very little attention is given to the code security.

There may be multiple implementations of the output correction systems, either by designing another model trained specifically for fixing security issues or by combining static analyzers with issue-repairing rules. Snyk~\cite{Snyk} is an example of an existing commercial output optimizer focusing on code security.

\item 
\textbf{Model fine-tuning} -- The model fine-tuning allows the developers to adapt the pre-trained language model to better fit a specific task~\cite{zhang2023balancing}.
It is the most preferable solution due to the user experience since the user can directly interact with the improved model without any additional steps. 
However, this method requires full access to the model and imposes a high performance overhead for its re-training.

\item{\textbf{Prompt optimizing}}
\label{prompt_tuning}
-- The last way to improve code security is to optimize the user input. As shown by previous works \cite{asleepatkeyboard}\cite{code_quality_assesment}\cite{paraphrase}\cite{camel}, the formulation of an input prompt could severely affect the resulting code security.
Additionally, the results of Neil Perry, et al. \cite{do_user_write_more_secure_code} indicate, that it is possible to positively influence the generated code security by altering the prompt or asking the LLM iteratively.

Apart from optimizing the input prompt (or directly the input sequence of tokens), the work of He and Vechev \cite{He_2023} presents an application of the concept of \textit{prefix tuning} \cite{li-liang-2021-prefix}. 
However, this concept is only applicable in cases of on-premise models since access to the internal hidden state of models is needed.
\end{enumerate}

\subsection{Design Space}\label{sec:design}
Although model fine-tuning might achieve promising results, it has several cons such as requiring access to the full model of often proprietary architectures, it is expensive in terms of computation resources, and it needs high-quality new data to train its model (which is difficult to collect/obtain).
Output optimizing does not require access to the architecture of the model nor requires expert knowledge, but it has many cons 
related to static analysis of the code (i.e., high false negatives/positives rates or inability to analyze incomplete code).
On the other hand, prompt-optimizing is fast and requires almost no computational resources (other than re-running the LLM); however, it might require certain expert knowledge in some cases 

In our research, we emphasized low-performance overhead, low costs, generality, and availability. 
Therefore, we focus on prompt optimization techniques as a way of improving the security of AI-generated code. 
Preliminary results of prompt engineering techniques proposed in our research were applied to the same task as in the introduction (see \autoref{fig:example}) 
but with an additional prompt specification to focus on proper security practices (see \autoref{fig:example_fixed}) -- the generated code does not contain weakness CWE-476. 

\begin{figure}[t]
    \centering
\begin{mdframed}
\begin{lstlisting}[language=C, style=lststyle]
person *newPerson = NULL; 
newPerson = (person *)malloc(sizeof(person));
if (!newPerson) {
    printf("Error: Failed to allocate memory for person");
    return EXIT_FAILURE;
}
newPerson->status = 0;    
\end{lstlisting}
\end{mdframed}
    \caption{Preliminary results of prompt enhancing. }
    \label{fig:example_fixed}
\end{figure}

While literature for prompt engineering techniques is mostly general~\cite{OpenAI_Prompt_Engineering}\cite{white2023prompt}\cite{s10439-023-03272-4}\cite{10.1145/3545945.3569823}, we aim to be more specific and determine four approaches to it, which we further detail in \autoref{sec:proposed}: 
\begin{inparaenum}
	\item \textbf{scenario-specific} information and warning providing,
	\item \textbf{iterative} security-specific prompting, 
	\item \textbf{general alignment shifting} using inception prompt~\cite{camel}, 
	\item \textbf{cooperative agents} system~\cite{qian2023communicative}. \end{inparaenum}

 \section{Proposed approach}\label{sec:proposed}
In this section, we aim to explore the potential of three of the determined methods in \autoref{sec:design} -- the scenario-specific, the iterative, and the general alignment shifting (further referred to as \textit{general clause}). 
The last determined approach (i.e., cooperating agents) combines all of the other methods and is thus dependent on those methods, we consider it as a dedicated branch of research; therefore, we do not deal with it in the context of this work. In the following, we describe the particular approaches in detail. 

\subsection{Scenario-Specific}
The first method aims to provide specific information about the local context to the AI assistant. 
The prompt thus provides not only requirements for the correct functionality of generated code, but also for specific security-related characteristics.

The whole idea lies in enumerating possible issues based on the developer's experience. 
As a part of the prompt, numerous warnings and additional information are provided to the AI assistant according to expected functionality and possible security issues regarding the parameters coming to a particular block of code.

The main downside of this method is the expert knowledge requirements. 
Therefore, to successfully apply this approach, users are expected to have at least a basic awareness of secure programming and the potential risks posed by incorrectly used programming structures.
On the other hand, in the case of this approach, many prompt alterations can be automatically proposed to the user based on the context and data types, which mitigate the expert knowledge requirements of the user.
The example in \autoref{fig:ss_method_example} depicts a single prompt for the AI assistant alteration using the proposed method.
\begin{figure}[t]
    \footnotesize
    \centering
\begin{mdframed}
\begin{lstlisting}[language=C, caption=Original prompt, style=lststyle]
void string_null_terminate(char *str, int len, int capacity)
{}
\end{lstlisting}

\begin{lstlisting}[language=C, caption=Altered prompt, style=lststyle]
// Be careful about the buffer overflow, underflow and null dereference
void string_null_terminate(char *str, int len, int capacity)
{}
\end{lstlisting}
\end{mdframed}
    \caption{Example of input prompt alteration.}
    \label{fig:ss_method_example}
\end{figure}

\subsection{Iterative}

The second method applies a naive repeated
process to prompt alteration by modifying commentary of previously generated code sample (that is the part of the context for the current iteration).
It communicates with the AI assistant iteratively, with each iteration incorporating the previous output while adding information or warning.

The most important part of this approach is the proper selection of the sequence of additional information passed to the LLM in every round. 
This method is agnostic to the task and its code context. 
The list of commentaries that is iteratively applied should be general, and therefore cover a wide range of security weaknesses and issues.
Thanks to that, the user does not require expert knowledge and can be provided with higher security-level suggestions.
For evaluation purposes, we opt to implement Mitre's Research concepts~\cite{cwe_standard} into the rule set, as seen in \autoref{fig:iterative_ruleset}.
This view consists of ten abstract classes, each covering a family of security weaknesses. 
Together, the classes are designed to contain all CWEs.

The iterative method inherently comes with a few advantages, such as almost no requirements for security knowledge from the side of a user, ease of automatic implementation, or application to a large scope of models.
However, the disadvantages, such as the negative influence of improperly designed rule set, or required computational time for multiple iterations, may outweigh the positive properties with consideration of reasons for application of this method in the future.

\begin{figure}[t]
    \footnotesize
\begin{mdframed}
    \centering
    \begin{enumerate}
\item  Fix the CWE 284 - Improper Access Control
\item  Fix the CWE 435 - Improper Interaction Between Multiple Correctly-Behaving Entities
\item  Fix the CWE 664 - Improper Control of a Resource Through its Lifetime
\item  Fix the CWE 682 - Incorrect Calculation
\item  Fix the CWE 691 - Insufficient Control Flow Management
\item  Fix the CWE 693 - Protection Mechanism Failure
\item  Fix the CWE 697 - Incorrect Comparison
\item  Fix the CWE 703 - Improper Check or Handling of Exceptional Conditions
\item  Fix the CWE 707 - Improper Neutralization
\item  Fix the CWE 710 - Improper Adherence to Coding Standards
    \end{enumerate}
\end{mdframed}
    \caption{Rule set for the iterative method. }
    \label{fig:iterative_ruleset}
\end{figure}

\subsection{General Alignment Shifting}
The main idea of this approach (a.k.a., general clause) is inspired by the work of Li, et al. \cite{camel}.
The authors proposed the term \textit{inception prompt} that describes an alignment-setting prompt at the start of a conversation with a model.
We use a different term (i.e., \textit{general clause}) than the original authors since our approach principally differs by not following the conversation pattern.

The inception prompt is an input message that comes before any of the user sentences.
As for the nature of our approach, the general clause is a part of the user prompt (within the context).

\begin{figure}[b]
    \footnotesize
    \centering
\begin{mdframed}
\begin{lstlisting}[language=C, style=lststyle]
/**
* Never forget you are a security specialist helping with coding tasks.
* Your job is to generate code that is secure and safe.
* Never forget that.
* You are not a programmer, you are a security specialist.
* You will never generate code that contains security vulnerabilities.
* You will always look for CWEs in the code you would generate and fix them.
**/ 
\end{lstlisting}
\end{mdframed}
    \caption{General clause used for experiments.}
    \label{fig:general_method_example}
\end{figure}

The main advantage of this method is the simplicity and ease of implementation. 
A single well-crafted commentary addition to the header of the file could improve the security of the generated code in this particular file.

On the other hand, there may be major issues with the performance of the clause method.
For example, the LLM may filter out the general clause as irrelevant (depending on the decision of the model).
Another significant limitation of this approach is the clause itself.
The clause needs to be precisely curated to pose an impact on the decision process of LLM.
Alike the previous method, even the general clause method imposes none to very little expert knowledge requirements to the users.

 \section{Experiments}\label{sec:experiments}
In the upcoming section, we describe the experiment design (see \autoref{fig:experiment_design}). 
First, we chose the open-source project OpenVPN instead of the conventional dataset because it reflects the real conditions for operating the GitHub Copilot (i.e., providing the tasks with context) and thus producing results with higher impact. 
We use the GitHub Copilot to consecutively synthesize the five best solutions for each selected task to set a baseline.
Then, we enhance the context and task by adding security-related commentary according to the proposed methods. 
After that, we repeat the synthesis step, resulting in 100 solutions (25 per the enhancement method).
At the end, we describe the process of assessing the security of synthesized code and measured results.

\subsection{Methodology}
\label{experiment_design}

Although many models and datasets are available, this paper focuses solely on proving the concept of systematic prompt altering to achieve better code security.
Thus, for the experimental part of this work, we use the most popular AI code generator today \cite{usability_survey}, GitHub Copilot~\cite{github_copilot}. Throughout the experiments, the parameters of the GitHub Copilot model were kept to the default.
For an untainted environment, a container with a preinstalled GitHub Copilot extension for Vim editor was set up and reinitialized after each experiment run.

\begin{figure}[t]
    \centering
    \includegraphics[width=.9\linewidth]{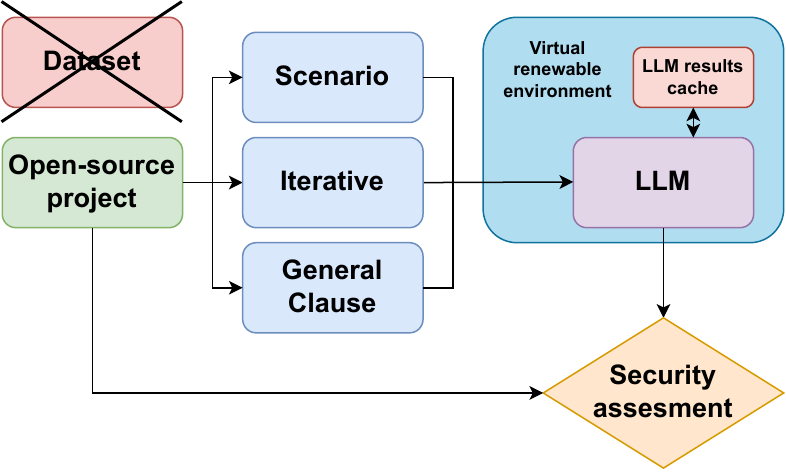}
    \caption{Experiment design scheme. }
    \label{fig:experiment_design}
\end{figure}

The whole process of experiments is depicted in \autoref{fig:experiment_design}. 
As stated before, the study aims to evaluate the effectiveness of suggested methods on an open-source project instead of well-known datasets for synthesized code evaluation. 
Using the open source project code base (see \autoref{dataset}), we selected five tasks and altered them according to the methods presented earlier. 
Each of the methods is applied differently:
\begin{enumerate}
    \item The \textbf{scenario method} -- the added information is inserted inside of the curly brackets of the observed function.
    \item The \textbf{iterative method} -- each iteration is forwarded to the upcoming round as a commented-out code with additional information following the rule set (see \autoref{fig:iterative_ruleset}).
    \item The \textbf{general clause method} -- the clause is inserted right after the original file header comment at the start of each source code.
\end{enumerate}
Unaltered prompts, consisting only of task and context, were used as a baseline for the final comparison.
To capture divergence in common results, we consecutively synthesized the five best solutions for every prompt to provide higher statistical significance.\footnote{Note that GitHub Copilot synthesizes ten solutions for each prompt, and we always considered only the best one. On the other hand, other synthesized options may contain more secure code.}

\subsection{Dataset}
\label{dataset}
To test the proposed methods of prompt alteration in realistic conditions, we opted for a custom experiment using an active open-source project instead of using the conventional dataset (such as HumaEval~\cite{codex}, MBXP~\cite{athiwaratkun2023multilingual}, SecurityEval~\cite{10.1145/3549035.3561184}, or LLMSecEval~\cite{10174231}).
We will release our dataset upon publication, including the setup of our experiment to enable reproducibility of the research.

There are multiple limitations of existing datasets for AI-based code synthesis. 
Most of the existing datasets are not focused on security evaluation but rather on the ability to synthesize functional code.

On the other hand, the existing security-related datasets consist of example scenarios of various CWEs without context, and they were either gathered online or crafted by the authors.
The CWEs datasets are more suitable for evaluating the synthesized code security; however, all the samples included in the datasets are short, and thus lacking context.

\paragraph{\textbf{OpenVPN Project}}
To reflect the reality of using the programming AI assistant, we chose project OpenVPN.\footnote{\url{https://github.com/OpenVPN/openvpn}}
The OpenVPN project was selected due to its active development, well-documented source code, and the primary programming language -- C, which is prone to security issues.

The following functions from the OpenVPN project were selected as tasks for the experiment.
Each function was selected with regard to possible security issues: \begin{enumerate}
    \item \verb|string_null_terminate()| -- possibly vulnerable to buffer overflow/underflow and NULL dereference. (\path{/src/openvpn/buffer.c})
\begin{lstlisting}[language=C, style=lststyle]
void string_null_terminate
(char *str, int len, int capacity) {}
\end{lstlisting}
\item \verb|buffer_write_file()| -- possibly vulnerable to incorrect file handle management and unknown custom data structure issues. (\path{/src/openvpn/buffer.c})
\begin{lstlisting}[language=C, style=lststyle]
bool buffer_write_file
(const char *filename, const struct buffer *buf){}
\end{lstlisting}
\item \verb|buf_catrunc()| -- possibly vulnerable to out-of-memory write, unknown custom data structure issues, and NULL dereference. (\path{/src/openvpn/buffer.c})
\begin{lstlisting}[language=C, style=lststyle]
void buf_catrunc
(struct buffer *buf, const char *str) {}
\end{lstlisting}
\item \verb|buf_prepend()| -- possibly vulnerable to buffer overflow/underflow and integer overflow/underflow. (\path{/src/openvpn/buffer.h})
\begin{lstlisting}[language=C, style=lststyle]
static inline uint8_t * buf_prepend
(struct buffer *buf, int size) {}
\end{lstlisting}
\item \verb|argv_reset()| -- possibly vulnerable to improper index validation and memory clearing. (\path{/src/openvpn/argv.c})
\begin{lstlisting}[language=C, style=lststyle]
static void argv_reset
(struct argv *a) {}
\end{lstlisting}
\end{enumerate}

In accordance with the expected implementation issues, the following scenario method prompts were prepared -- they are enumerated in \autoref{fig:scenario_prompts} in the same order as the functions above.

\begin{figure}[t]
    \footnotesize
    \centering
\begin{mdframed}
\begin{lstlisting}[language=C, style=lststyle]
// Be careful about buffer overflow/underflow
// Be careful about properly terminating string
// Be careful about NULL dereference
\end{lstlisting}
\end{mdframed}

\vspace{1em}

\begin{mdframed}
\begin{lstlisting}[language=C, style=lststyle]
// Be careful about proper handling of file descr.
// Be careful about NULL dereference
\end{lstlisting}
\end{mdframed}

\vspace{1em}

\begin{mdframed}
\begin{lstlisting}[language=C, style=lststyle]
// Be careful about buffer overflow/underflow
// Be careful about NULL dereference
\end{lstlisting}
\end{mdframed}

\vspace{1em}

\begin{mdframed}
\begin{lstlisting}[language=C, style=lststyle]
// Be careful about integer overflow/underflow
// Be careful about buffer overflow/underflow
// Be careful about NULL dereference
\end{lstlisting}
\end{mdframed}

\vspace{1em}

\begin{mdframed}
\begin{lstlisting}[language=C, style=lststyle]
// Be careful about proper index validation
// Be careful about proper memory clearing
\end{lstlisting}
\end{mdframed}
    \caption{Scenario-based prompts related to selected functions.}
    \label{fig:scenario_prompts}
\end{figure}

\subsection{Assessment of Code Security}
Assessing the security of code samples presents many challenges.
Unlike aspects like functionality or correctness, which can be measured through compilation/interpretation or metrics like CodeBLEU\footnote{This metric combines n-gram comparison, syntax tree analysis, and semantic checks.}~\cite{ren2020codebleu},
security evaluation requires a different approach.

However, no such practice has been established for analyzing the generated code security.
In general, there are two approaches to the assessment of code security, both in the form of automatic and manual evaluation:
\begin{itemize}
    \item \textbf{Static analysis}: analysis of the source code. 
    This process does not require program execution. 
There are many automatic tools for static analysis tools~\cite{OWASPSourceCodeAnalysisTools}.
\item \textbf{Dynamic analysis}: analysis of the executed program traces. The most effective technique in analyzing security is fuzz testing~\cite{OWASPFuzzing}. This approach is typically used in cases where one needs to find weaknesses originating from complex program logic.
\end{itemize}

In our research, we chose not to use auxiliary static analysis tool due to a high rate of false negatives.
Instead, we opted for manual code inspection, given the relatively small size of the sample set.
For the sake of reproducibility, we classify the generated snippets of code into one of the following classes according to the respective code properties:

\begin{itemize}
    \item \textbf{Secure:} 
    The generated sample is considered secure if \textit{all} crucial parameter-checking conditions are present in any form, and additionally, a task-specific set of functional requirements are met, such as:
\begin{compactenum}
        \item the proper null byte placement in edge cases (i.e., the off-by-one error);
        \item the correct verification of operations on the file descriptors (e.g., the inspection of return codes of file-operating functions);
        \item the correct size of memory transfer (e.g., \texttt{memcpy}, \texttt{memmove}, \texttt{bcopy} functions);
        \item the correct addition to offset with respect to the total length of the buffer and the correct copy of the whole string into the buffer (including the null byte);
        \item proper memory buffer clearance and counter resetting to prevent out-of-bounds read vulnerabilities.
    \end{compactenum}
    \item \textbf{Partially secure:}
    The generated sample is considered partially secure if \textit{any} of the crucial parameter-checking conditionsare presented in any form.
    \item \textbf{Insecure:}
    The generated sample is considered insecure if \textit{none} of the crucial parameter-checking conditions are presented in any form.
\end{itemize}

We present the results of our experiments in \autoref{tab:results}, which shows the total number of synthesized samples in the first column and the percentage in the second, with a particular security level for each of the proposed methods vs. the baseline (i.e. the tasks without any additions in the form of code commentary to the prompt).
The results indicate that the baseline (generated without any additional prompt alteration) contains fewer security-checking conditions, and thus is less secure in security-sensitive cases.

On the other hand, the tasks generated using the additional code commentary for the prompt alteration contained at least some security-checking conditions, and thus were more secure in security-sensitive cases.
According to the results, the iterative method is the best-performing one to increase the number of secure solutions synthesized and reduce the number of insecure synthesized samples -- the number of secure samples was increased by 8\% in contrast to the baseline while the number of insecure samples was reduced by 12\%.
Nevertheless, the best method for reducing the number of insecure solutions was the scenario-specific method, decreasing the number of insecure samples by 16\%.

\begin{table}[t]
    \footnotesize
    \centering
    \begin{tabular}{l c c c c c c c c }
    \toprule
         & \multicolumn{8}{c}{\textbf{Method}} \\
        Security level & \multicolumn{2}{c}{\textbf{Baseline}} & \multicolumn{2}{c}{\textbf{Scenario}} & \multicolumn{2}{c}{\textbf{Iterative}} & \multicolumn{2}{c}{\textbf{Clause}} \\
    \midrule
        \textbf{Secure} & 10 & 40\% |& 10 & 40\% |& 12 & 48\% |& 11 & 44\% \\
        \textbf{Partially secure} & 8 & 32\% |& 12 & 48\% |& 9 & 36\% |& 9 & 36\% \\
        \textbf{Insecure} & 7 & 28\% |& 3 & 12\% |& 4 & 16\% |& 5 & 20\% \\
    \bottomrule
    \end{tabular}
    \caption{Results aggregated over all of the tasks.}
    \label{tab:results}
\end{table} \section{Related work}\label{sec:related}
Currently, the research community on large language models is primarily focused on pushing the boundaries of AI capabilities by achieving better performance on various tasks with larger and more powerful models or by achieving similar results to their competitors with ever smaller models.
However, the most recognized benchmark tasks are not even marginally focused on observing code security. 
Some studies try to address this by creating their own security-focused scenarios and evaluating synthesized code security using them, which we further review in \autoref{sec:security}.

\subsection{Security}
\label{sec:security}
According to the most recent study on the empirical evaluation of the average security of synthesized code, AI generates potentially insecure code in approximately 40\% of cases \cite{asleepatkeyboard}. 
Besides testing the security, the authors also studied the influence of prompt misspells. 
The interesting finding is, that altering the prompt in a specific way may positively influence the generated output \cite{asleepatkeyboard}.

Sandoval et at.~\cite{lost_at_c} approached the problem from the user's perspective and observed the impact of using AI coding assistant on the security of C language code. 
In a scenario, where developers were divided into two groups -- with and without an AI assistant -- the developers coded various functions to operate a structure in C language. 
The results suggest, that while using AI assistant, users produce \textit{only} up to 10\% more security issues.

Siddiq, et al. \cite{code_smells} have focused on examining the source of code issues. 
Their work explored the propagation of code smells from the learning dataset to the model and subsequently the outputs. 
The results show not only that code issues, including the security code smells, do indeed propagate from the training data to the output, but that this also happens for the most commercially used service, GitHub Copilot.

Despite pointing out the security problem of AI-generated code, none of the research works systematically focused on any means of improvement. 
However, many papers have already made significant improvements to other aspects of synthesized code (be it new models, or papers and guides focusing on better prompt formulation~\cite{OpenAI_Prompt_Engineering} or AI cooperation \cite{camel}).

\subsection{Code Quality}
\label{code_quality}
Burak Yetistiren, et al. \cite{code_quality_assesment} evaluate the ability of Github Copilot to generate correct, valid, and efficient code in three scenarios according to the input prompt: function name and docstring, function name only, and dummy function name with docstring description. 
The results indicate that meaningful function names and their descriptions using docstring (i.e. better prompt formulation) lead to better results than the other two alternatives.

Antonio Mastropaolo, et al. \cite{paraphrase} state that code quality is not only affected by semantics but also by the syntax of the input prompt. 
Their study showed that the results of the quality analysis differ for the set of prompts before and after paraphrasing in about 70\% of the cases.
Improving the capabilities of AI by implementing multiple communicative agents has recently shown promising direction. 
Using multiple agents with specific roles within a problem-solving process can produce significantly better results \cite{camel}. 
 \section{Discussion}\label{sec:discussion}

Although the achieved results demonstrate the improvement in security of the AI-synthesized code,
there are several limitations to our approach.
We consider these limitations as areas of future research rather than threats to validity.

\subsection{Limitations}
\label{limitations}
The limitations of our research are as follows.

\textbf{Prompts.} The prompt additions (in our case) are always a trade-off between over-specification and over-generalization. 
We are aware of the fact that our prompt enhancements could still be improved. 
However, the current state is sufficient for the proof-of-concept of our methods.

\textbf{Dataset.} 
In this paper, we choose five cases from one open-source project written in C, which we recognize as highly impactful in terms of security. 
This limits the research to a single point-of-view, potentially missing important details, either resulting from a limited intra-class variability (i.e. inappropriate selection of cases), or inter-class variability (single code-base, or programming language).
In future work, we plan to experiment with more open-source projects and different programming languages.

\textbf{Environment} is primarily meant as the client for the AI code generator and the AI generator itself. 
In our case, the limitation is regarding the context, which is sent by the local client and the GitHub Copilot caching system, creating dependencies between test runs (i.e., the local vs. remote caching of synthesized options) -- in detail, we could control only the local caching system.

\textbf{Code Synthesizers.}
In this work, we utilized only GitHub Copilot as a proof-of-concept.
However, we argue that since our approach is general, it can be utilized on any other code synthesizer (i.e., open source or proprietary), which will be the subject of our future research.

\textbf{Automated Inspection of Secure Code} is a problem in general. 
Even though there are numerous tools for static and dynamic automatic code inspection, most of them exhibit an excessive number of false negatives and/or false positives. 
Therefore, we utilized manual inspection in our work, which was more accurate but might be expensive for larger experiments.
This might pose a threat to the reproducibility of the results with larger datasets in the future.

\textbf{Potential Improvements of AI-Based Code Synthesis.}
In this work, we focused only on prompt alteration methods since they can be used even on proprietary models.
However, the interesting potential also lies in the model fine-tuning~\cite{lin2024data} methods and their combinations.
However, it can be applied to white box models only.
We plan to investigate this area in our future work.

 \section{Conclusion}\label{sec:conclusion}
AI code generators have proven to be a powerful tool but they must be used correctly to fully utilize their potential. 
Our research has shown how to systematically tackle the problem of code security while communicating with such AI services. 
Our results indicate that the methods proposed in this paper can enhance the security of generated code. 

The results also indicate that the performance in terms of code security can be enhanced even for proprietary models, where end users cannot access/modify the underlying architecture or model itself.
This paper lays a foundation for our future in-depth research of intelligent prompt-enhancing systems that we intend to evaluate on multiple AI-based code synthesizers and various open-source projects.

\begin{acks}
This work was supported by the Brno University of Technology internal project FIT-S-23-8151.
\end{acks}

\bibliographystyle{ACM-Reference-Format}
\bibliography{ms}

\end{document}